\documentclass[10pt]{article}
\usepackage[square,authoryear]{natbib}
\usepackage{graphicx}
\usepackage{stackrel}
\usepackage{cancel}

\usepackage[all]{xy}

\usepackage{epstopdf,framed}
\usepackage{pdfsync}
\usepackage{amscd}
\usepackage{mathrsfs}

\RequirePackage{color}
\usepackage{amsmath,bm}
\usepackage{amssymb}
\usepackage{amsfonts}
\usepackage{url}

\begin{document}
\newtheorem{remark}[theorem]{Remark}

\title{From variational to bracket formulations in nonequilibrium thermodynamics of simple systems}
\vspace{-0.2in}

\newcommand{\todoFGB}[1]{\vspace{5 mm}\par \noindent
\framebox{\begin{minipage}[c]{0.95 \textwidth} \color{red}FGB: \tt #1
\end{minipage}}\vspace{5 mm}\par}

\date{}
\author{\hspace{-1cm}
\begin{tabular}{cc}
Fran\c{c}ois Gay-Balmaz &
Hiroaki Yoshimura 
\\  CNRS, LMD, IPSL & School of Science and Engineering
\\  Ecole Normale Sup\'erieure & Waseda University
\\  24 Rue Lhomond 75005 Paris, France & Okubo, Shinjuku, Tokyo 169-8555, Japan
\\ gaybalma@lmd.ens.fr & yoshimura@waseda.jp\\
\end{tabular}\\\\
}

\maketitle
\vspace{-0.3in}

\begin{center}
\abstract{A variational formulation for nonequilibrium thermodynamics was recently proposed in \cite{GBYo2017a,GBYo2017b} for both discrete and continuum systems. This formulation extends the Hamilton principle of classical mechanics to include irreversible processes.  In this paper, we show that this variational formulation yields a constructive and systematic way to derive from a unified perspective several bracket formulations for nonequilibrium thermodynamics proposed earlier in the literature, such as the single generator bracket and the double generator bracket. In the case of a linear relation between the thermodynamic fluxes and the thermodynamic forces, the metriplectic or GENERIC bracket is recovered.
We also show how the processes of reduction by symmetry can be applied to these brackets. In the reduced setting, we also consider the case in which the coadjoint orbits are preserved and explain the link with double bracket dissipation.
A similar development has been presented for continuum systems in \cite{ElGB2019} and applied to multicomponent fluids.
}
\end{center}

\tableofcontents

\section{Introduction}

A Lagrangian variational formulation for nonequilibrium thermodynamic has been proposed in the papers \cite{GBYo2017a,GBYo2017b} for finite dimensional and continuum closed systems and for open systems in \cite{GBYo2018a}. This variational formulation extends the Hamilton principle of classical mechanics to include irreversible processes such as friction, heat or mass transfer in the equations of motion. It is a type of \textit{Lagrange-d'Alembert principle} with nonlinear constraints and it follows a very systematic construction from the given thermodynamic fluxes and forces of the irreversible processes. This formulation is based on the concept of \textit{thermodynamic displacements} which are defined as the primitive in time of the thermodynamic forces. This variational formulation has a naturally associated geometric description given in terms of \textit{Dirac structures}, as shown in \cite{GBYo2018b}.

Historically, the  proposed general formalisms for nonequilibrium thermodynamics have been mainly constructed via appropriate \textit{modifications of Poisson brackets}, as initiated by \cite{Ka1984,Mo1984a,Gr1984}. Since then, this approach has been developed for a large list of systems, see, e.g. \cite{GrOt1997}. Other classes of brackets have been proposed, e.g. \cite{BeEd1994,EdBe1991a,EdBe1991b}.
Unlike the variational formalism, most of these bracket formalisms do not follow from a systematic construction but have often been derived via a case-by-case approach, with slightly different axioms used in different situations.

In this paper, we show that the variational formulation systematically yields the two main bracket formalisms, namely, the single and double generator brackets. Moreover, in the case of a linear relation between the thermodynamic fluxes and the thermodynamic forces,
the metriplectic (\cite{Mo1986}) or GENERIC (\cite{GrOt1997,OtGr1997}) bracket is recovered. Specifically, we focus on the case of simple thermodynamic systems, in which only one entropy variable is needed, but allowing for internal mass transfer. The general case will be studied elsewhere.
We also consider the  reduced versions of these brackets for systems on Lie groups, by using the reduction by symmetry of the variational formulation of thermodynamics developed in \cite{CoGB2019}.

The derivation of such brackets from the variational formulation for continuum system has been illustrated in \cite{ElGB2019} in the context of multicomponent fluids.

\section{Variational formulation of nonequilibrium thermodynamics}

In this section we review from \cite{GBYo2017a} the variational formulation for the thermodynamics of adiabatically closed and simple systems. We start with the simplest case of mechanical systems with friction and then extend it to the case with internal mass transfer.

\subsection{Variational formulation for mechanical systems with friction}

Consider a thermodynamic system described only by a mechanical variable $q\in Q$ and an entropy variable $S\in \mathbb{R}$. The Lagrangian of this thermodynamic system is a function
\[
L: TQ \times \mathbb{R}  \rightarrow \mathbb{R} , \quad (q, v, S) \mapsto L(q, v, S),
\]
where $TQ$ denotes the tangent bundle of the mechanical configuration manifold $Q$.
We assume that the system involves external and friction forces given by fiber preserving maps $F^{\rm ext}, F^{\rm fr}:TQ\times \mathbb{R} \rightarrow T^* Q$, i.e., such that $F^{\rm fr}(q, v, S)\in T^*_qQ$, similarly for $F^{\rm ext}$. As stated in \cite{GBYo2017a}, the variational formulation for this system is given as follows:

\medskip

\begin{framed}
\noindent Find the curves $q(t)$, $S(t)$ which are critical for the \textit{variational condition}
\begin{equation}\label{LdA_thermo_simple}
\begin{aligned}
&\delta \int_{t _1 }^{ t _2}L(q , \dot q , S){\rm d}t +\int_{t_1}^{t_2} \left<F^{\rm ext}(q, \dot q, S), \delta q \right>\,{\rm d}t =0\,,
\end{aligned}
\end{equation}
subject to the \textit{phenomenological constraint}
\begin{equation}\label{CK_simple} 
\frac{\partial L}{\partial S}(q, \dot q, S)\dot S  =  \left<F^{\rm fr}(q, \dot q, S), \dot q \right>,\qquad 
\end{equation}
and for variations subject to the  \textit{variational constraint}
\begin{equation}\label{CV_simple} 
\frac{\partial L}{\partial S}(q, \dot q, S)\delta S=  \left<F^{\rm fr}(q , \dot q , S), \delta q \right>,
\end{equation}
with $ \delta q(t_1)=\delta q(t_2)=0$.\vspace{-0.0cm}
\end{framed}
This variational formulation yields the system of equations
\begin{equation}\label{simple_systems} 
\frac{d}{dt}\frac{\partial L}{\partial \dot q}-\frac{\partial L}{\partial q}= F^{\rm fr}(q,\dot q, S),\qquad \frac{\partial L}{\partial S}\dot S= \langle  F^{\rm fr}(q,\dot q, S), \dot q \rangle.
\end{equation}
The first equation is the balance of mechanical momentum, while the second one gives the rate of entropy production of the system 
\[
\dot S= -\frac{1}{T}\left<F^{\rm fr}(q,\dot q, S),\dot q\right>,
\]
with $T=-\frac{\partial L}{\partial S}(q, \dot q, S)$ the temperature of the system. From the second law the friction force $F^{\rm fr}$ must satisfy
\begin{equation}\label{condition_F_fr}
\left<F^{\rm fr}(q, \dot q, S),\dot q \right>\leq 0,\;\; \text{for all}\;\;(q,\dot q,S).
\end{equation}
For instance, for a friction force linear in velocities, we have
\[
F^{\rm fr}_{i}=-\lambda_{ij} \dot{q}^{j},
\]
where $\lambda_{ij}$, $i,j=1,...,n$ are functions of the state variables with the symmetric part of the matrix $\lambda_{ij}$ positive semi-definite.

\subsection{Variational formulation for systems with internal mass transfer}

The previous variational formulation can be extended to systems experiencing internal diffusion processes. Diffusion is particularly important in biology, where many processes depend on the transport of chemical species through bodies, \cite{OsPeKa1973}. Consider a thermodynamic system consisting of $K$ compartments that can exchange matter by diffusion across walls (or membranes) on their common boundaries. We assume that the system has a single species and denote by $N _k $ the number of moles of the species in the $k$-th compartment, $k=1,...,K$. We assume that the thermodynamic system is simple; i.e., a uniform entropy $S$, the entropy of the system, is attributed to all the compartments. The Lagrangian of this thermodynamic system is thus a function
\begin{equation}\label{L_diffusion}
\begin{aligned}
L: &\,TQ \times \mathbb{R}^{K+1}  \rightarrow \mathbb{R},\;\;
(q, v, S, N_1,..,N_K) \mapsto L(q, v, S, N_1,..,N_K).
\end{aligned}
\end{equation}
We denote $\mathcal{J} ^{\ell \rightarrow k}=- \mathcal{J} ^{k \rightarrow \ell}$ the molar flow rate from compartment $\ell$ to compartment $k$ due to diffusion of the species. In general, we have the dependence
\begin{equation}\label{dependence_Jkl}
\mathcal{J} ^{\ell \rightarrow k}=\mathcal{J} ^{\ell \rightarrow k}\left(S, N_k,N_\ell, \frac{\partial L}{\partial N_k},\frac{\partial L}{\partial N_\ell}\right).
\end{equation}

The variational formulation involves the new variables $W^k$, $k=1,...,K$, which are examples of \textit{thermodynamic displacements} and play a central role in our formulation. In general, we define the \textit{thermodynamic displacement associated to an irreversible process} as the primitive in time of the thermodynamic force (or affinity) of the process. This force (or affinity) thus becomes the rate of change of the thermodynamic displacement.
In the case of matter transfer, $\dot W^k$ corresponds to the chemical potential of $N_k$.
The variational formulation for a simple system with internal diffusion process is stated as follows.

\medskip

\begin{framed}
\noindent Find the curves $q(t)$, $S(t)$, $W^k(t)$, $N_k(t)$ which are critical for the \textit{variational condition}
\begin{equation}\label{VCond_simple_diffusion}
\begin{aligned}
&\delta \int_{t _1 }^{ t _2} \!\Big[ L\left(q, \dot q, S, N_1,...,N_K\right)+ \dot W^kN_k\Big] {\rm d}t +\int_{t_1}^{t_2} \left<\!F^{\rm ext }, \delta q\right>\,{\rm d}t=0, 
\end{aligned}
\end{equation}
subject to the \textit{phenomenological constraint}
\begin{equation}\label{PC_simple_diffusion}
\frac{\partial L}{\partial S}\dot S  =  \left<F^{\rm fr}, \dot q\right>   + \sum_{k,\ell=1}^{K}  \mathcal{J}^{\ell \rightarrow k} \dot W^k,
\end{equation}
and for variations subject to the \textit{variational constraint}
\begin{equation}\label{VC_simple_diffusion}
\frac{\partial L}{\partial S}\delta S  = \left< F^{\rm fr},\delta q \right> + \sum_{k,\ell=1}^{K}  \mathcal{J}^{\ell \rightarrow k} \delta W ^k,
\end{equation}
with $\delta q(t_1)=\delta q(t_2)=0$ and $ \delta W^k(t_1)=\delta W^k(t_2)=0$, $k=1,...,K$.\vspace{-0.0cm}
\end{framed}

These conditions, combined with the phenomenological constraint \eqref{PC_simple_diffusion}, yield the following system of evolution equations for the curves $q(t)$, $S(t)$, and $N^k(t)$:
\begin{equation}\label{simple_systems_matter} 
\left\{
\begin{array}{l}
\displaystyle\vspace{0.2cm}\frac{d}{dt}\frac{\partial L}{\partial \dot q}- \frac{\partial L}{\partial q}=  F^{\rm fr} +F^{\rm ext} ,\\
\displaystyle\vspace{0.2cm}\frac{d}{dt} N _k = \sum_{\ell=1}^K \mathcal{J} ^{\ell \rightarrow k}, \quad k=1,...,K,\\
\displaystyle\frac{\partial L}{\partial S}\dot S=  \left<F^{\rm fr},\dot q \right> - \sum_{k<\ell} \mathcal{J}^{\ell\rightarrow k}\left(\frac{\partial L}{\partial N_k} - \frac{\partial L}{\partial N_\ell}\right) .
\end{array} \right.
\end{equation}
The last equation in \eqref{simple_systems_matter} yields the rate of entropy production of the system as
\begin{equation}\label{entropy_production_Nk}
\dot S=-\frac{1}{T} \left<F^{\rm fr}, \dot{q} \right> - \frac{1}{T}\sum_{k<\ell}
\mathcal{J} ^{\ell \rightarrow k} (\mu^k -\mu^\ell),
\end{equation}
with $\mu^k= -\frac{\partial L}{\partial N_k}$ the chemical potentials. The two terms in the right-hand side of \eqref{entropy_production_Nk} correspond, respectively, to the rate of entropy production due to mechanical friction and that due to matter transfer.
From the second law, $F^{\rm fr}$ and $\mathcal{J} ^{k \rightarrow \ell}$ must satisfy
\begin{equation}\label{thermo_consistency}
\left<F^{\rm fr}, \dot{q} \right>\leq 0 \qquad\text{and}\qquad \mathcal{J} ^{\ell \rightarrow k} (\mu^k -\mu^\ell)\leq 0.
\end{equation}
When a linear relation is assumed between the forces and fluxes, we have relations
\begin{equation}\label{thermo_consistency_linear}
F^{\rm fr}_{i}=-\lambda_{ij} \dot{q}^{j}\qquad\text{and}\qquad \mathcal{J} ^{\ell \rightarrow k}=-G^{k\ell} (\mu ^k-\mu^\ell),
\end{equation}
where $\lambda_{ij}$, $i,j=1,...,n$ and $G^{k\ell}$, $k,\ell=1,...,K$ are functions of the state variables, with the symmetric part of the matrix $\lambda_{ij}$ positive semi-definite and with $G^{k\ell}\geq 0$, for all $k,\ell$.

Note that in both variational formulations \eqref{LdA_thermo_simple}--\eqref{CV_simple} and \eqref{VCond_simple_diffusion}--\eqref{VC_simple_diffusion}, the two constraints are related in a very systematic way, suggested by the relation
\begin{equation}\label{relation_CV_CK}
\sum_\alpha J_\alpha\dot\Lambda^\alpha\;\; \leadsto\;\; \sum_\alpha J_\alpha\delta\Lambda^\alpha,
\end{equation}
with $J_\alpha$ the thermodynamic flux and $\Lambda^\alpha$ the thermodynamic displacement of the process $\alpha$. This systematic correspondence holds for finite dimensional and continuum closed systems, and is at the core of the formulation in terms of Dirac structures, \cite{GBYo2017a,GBYo2017b,GBYo2018b}.

For simplicity, from now on we set the external forces $F^{\rm ext}$ to zero. They can be easily included in our developments below, and yield an additional term in the various bracket formalisms.

\section{Single and double generator brackets}

In this section we shall show that the variational formulation has the property to systematically induce and unify several bracket formulations for nonequilibrium thermodynamics proposed earlier in the literature, such as the single generator bracket, the double generator bracket, and the metriplectic (or GENERIC) bracket.

\subsection{Bracket formulations in nonequilibrium thermodynamics}

There are two main approaches to the bracket formulation for irreversible processes in the literature: the \textit{single generator} and \textit{double generator} formulations. In this paragraph we quickly review the structure of these two brackets. Let $M$ be a Poisson manifold, with Poisson bracket $\{\,,\}$. We denote by $H\in C^\infty(M)$ the Hamiltonian and $S\in C^\infty(M)$ the entropy. We assume that $\{H,S\}=0$.

In the single generator formalism, \cite{BeEd1994,EdBe1991a,EdBe1991b}, the evolution of an arbitrary functional $F\in C^\infty(M) $ is governed by 
\begin{equation}\label{single_generator}
\frac{d}{dt}F = \{F ,H \} + [F ,H ],
\end{equation}
where the dissipation bracket $[F ,H ]$ is linear and a derivation in $F $, it can be nonlinear in $H $, and satisfies $[H ,H ] = 0$ and $[S ,H ] \geq 0$. These last two requirements are the first and second laws of thermodynamics, respectively. Since both the reversible (Poisson) and dissipation brackets use the same generator $H $, this is referred to as the \textit{single generator formalism}. The bracket formulation \eqref{single_generator} yields the dynamical system $\dot m(t)= X_H(m(t))+ D_H(m(t))$, where $X_H= J\mathbf{d}H$ is the Hamiltonian vector field associated to $H$, with $J:T^*M \rightarrow TM$ the Poisson tensor, and the vector field $D_H$ is determined from $[F,H]=\mathbf{d}F\cdot D_H$, for all $F$, which follows since $F\mapsto [F,H]$ is a derivation.

In the double generator formalism, the evolution of an arbitrary functional $F\in C^\infty(M) $ is governed by 
\begin{equation}\label{double_generator}
\frac{d}{dt} F = \{F ,H \} + (F ,S ),
\end{equation}
where the dissipation bracket $(F ,G )$ is symmetric, bilinear and satisfies the Leibniz rule, as well as $(H ,S ) = 0 $ and $(S ,S ) \geq 0$. These are precisely the axioms given in  \cite{Ka1984}. Since the Poisson and dissipation brackets use different generators ($H $ for Poisson and $S $ for dissipation), this is referred to as the \textit{double generator formalism}. The bracket formulation \eqref{double_generator} yields the dynamical system $\dot m(t)= J\mathbf{d}H(m(t))+ K\mathbf{d}S(m(t))$, where as before $J\mathbf{d}H=X_H$ is the Hamiltonian vector field associated to $H$, and the symmetric vector bundle linear map $K:T^*M\rightarrow TM$, $K^*=K$, is such that $(F,G)= \langle\mathbf{d}F, K\mathbf{d}G\rangle$, which follows from the fact that $(F,G)$ is symmetric and a derivation in each factor, and where $K^*: T^\ast M \to TM$ is the dual map of $K$, given by $\langle K^\ast \alpha, \beta\rangle= \langle\alpha, K\beta\rangle$, for all $\alpha, \beta \in T^\ast M$.

Sometimes, the stronger requirements that $\{G, S\}=0$, $(H ,G) = 0$, $(G, G)\geq 0$, for arbitrary $G\in C^\infty(M)$ are imposed, in which case the system \eqref{double_generator} is termed \textit{metriplectic}, \cite{Mo1986}. For example, this is what is used in the GENERIC formalism, see \cite{GrOt1997,OtGr1997}. When considering macroscopic systems, typically only bilinearity, $(H ,S ) = 0 $, and $(S ,S ) \geq 0$ seem to be required on physical grounds.

\subsection{Derivation of the single generator bracket}

Consider the system \eqref{simple_systems_matter}, assume that the Lagrangian $L$ in \eqref{L_diffusion} is hyperregular with respect to the mechanical part and define the associated Hamiltonian $H:T^*Q \times \mathbb{R}^{K+1}\rightarrow\mathbb{R}$ by
\[
H(q,p, S, N_1,...,N_K)= \langle p, v \rangle - L(q, v, S, N_1,...,N_K),
\]
where $v$ is such that $\frac{\partial L}{\partial v}=p$.
In terms of $H$, system \eqref{simple_systems_matter} can be equivalently written as
\begin{equation}\label{simple_systems_matter_H}
\hspace{-0.1cm}\left\{
\begin{array}{l}
\displaystyle\vspace{0.2cm} \dot q=\frac{\partial H}{\partial p}, \quad \dot p=-\frac{\partial H}{\partial q} + F^{\rm fr},\quad\frac{d}{dt} N _k = \sum_{\ell=1}^K \mathcal{J} ^{\ell \rightarrow k},\\
\displaystyle-\frac{\partial H}{\partial S}\dot S=  \Big\langle  F^{\rm fr}, \frac{\partial H}{\partial p}\Big \rangle + \sum_{k<\ell} \mathcal{J}^{\ell\rightarrow k}\Big(\frac{\partial H}{\partial N_k} - \frac{\partial H}{\partial N_\ell}\Big) .
\end{array} \right.
\end{equation}
In this system, the dependence of the fluxes in \eqref{dependence_Jkl} is written in terms of the Hamiltonian $H$ as
\begin{equation}\label{dependence_fluxes}
\begin{aligned}
F^{\rm fr}&= F^{\rm fr}\Big(q, \frac{\partial H}{\partial p}, S\Big),\\
\mathcal{J}^{\ell\rightarrow k}&= \mathcal{J}^{\ell\rightarrow k}\Big(S, N_k, \frac{\partial H}{\partial N_k},N_\ell, \frac{\partial H}{\partial N_\ell}\Big).
\end{aligned}
\end{equation}
For a given function $F\in  C^\infty(T^*Q\times \mathbb{R}^{K+1})$, by computing its time derivative
\[
\frac{d}{dt}F= \Big\langle \frac{\partial F}{\partial q}, \dot q\Big\rangle +  \Big\langle \frac{\partial F}{\partial p}, \dot p\Big\rangle +  \frac{\partial F}{\partial S} \dot S + \sum_{k=1}^K \frac{\partial F}{\partial N_k} \dot N_k,
\]
along a solution curve of \eqref{simple_systems_matter_H}, we directly deduce the form \eqref{single_generator}, with $\{\,,\}$ the direct sum of the canonical Poisson bracket on $T^*Q$ and the zero bracket on $\mathbb{R}^{K+1}$, where the dissipation bracket is computed as
\begin{equation}\label{bracked_EB}
\begin{aligned}
& [F ,H ]=\Big\langle  F^{\rm fr}, \frac{\partial F}{\partial p}\Big\rangle + \sum_{k<\ell} \mathcal{J}^{\ell\rightarrow k}\Big(\frac{\partial F}{\partial N_k} - \frac{\partial F}{\partial N_\ell}\Big)\\
&\qquad\qquad-\frac{\frac{\partial F}{\partial S}}{\frac{\partial H}{\partial S}}\Big[ \Big\langle   F^{\rm fr}, \frac{\partial H}{\partial p}\Big\rangle +\sum_{k<\ell} \mathcal{J}^{\ell\rightarrow k}\Big(\frac{\partial H}{\partial N_k} - \frac{\partial H}{\partial N_\ell}\Big)\Big].
\end{aligned}
\end{equation}
In this expression we recall that both $F^{\rm fr}$ and $\mathcal{J}^{\ell\rightarrow k}$ may depend on $H$ via \eqref{dependence_fluxes}. One directly checks that the conditions $\{H,S\}=0$, $[H,H]=0$ are satisfied. The condition $[S,H]\geq 0$ is satisfied if and only if \eqref{thermo_consistency} holds.

We have thus recovered the single generator formalism from the variational approach. This formulation does not impose a specific dependence (such as a linear dependence) of the thermodynamic fluxes\footnote{In the terminology of thermodynamics, the {\it friction force $F^{\rm fr}$ in mechanics} may be regarded as {\it thermodynamic flux} (not 'thermodynamic' force nor affinity) by convention.} $F^{\rm fr}$ and $\mathcal{J}^{\ell\rightarrow k}$ on the thermodynamic forces.

\subsection{Derivation of the double generator bracket}\label{derivation_double}

Starting again from the system \eqref{simple_systems_matter} obtained from the variational formulation, we compute as before the time derivative of an arbitrary function $F\in C^\infty(T^*Q\times \mathbb{R}^{K+1})$ along a solution of \eqref{simple_systems_matter_H}. The expression \eqref{bracked_EB} has now to be interpreted as the bracket $(F,S)$. Hence it suffices to multiply this expression by $1=\frac{\partial S}{\partial S}$, to symmetrize in $F$ and $S$ the resulting expression, and finally to replace $S$ by an arbitrary function $G$ to finally get the symmetric bracket
\begin{equation}\label{bracked_Kaufman}
\begin{aligned}
&(F,G)=\Big\langle  F^{\rm fr}, \frac{\partial F}{\partial p}\Big\rangle\frac{\partial G}{\partial S} + \Big\langle  F^{\rm fr}, \frac{\partial G}{\partial p}\Big\rangle\frac{\partial F}{\partial S}\\[3mm]
&\qquad\qquad+ \sum_{k<\ell} \mathcal{J}^{\ell\rightarrow k}\Big(\frac{\partial F}{\partial N_k} - \frac{\partial F}{\partial N_\ell}\Big)\frac{\partial G}{\partial S} + \sum_{k<\ell} \mathcal{J}^{\ell\rightarrow k}\Big(\frac{\partial G}{\partial N_k} - \frac{\partial G}{\partial N_\ell}\Big)\frac{\partial F}{\partial S}\\[3mm]
&\qquad\qquad-\frac{1}{\frac{\partial H}{\partial S}}\Big[ \Big\langle   F^{\rm fr}, \frac{\partial H}{\partial p}\Big\rangle \!+\!\sum_{k<\ell} \mathcal{J}^{\ell\rightarrow k}\Big(\frac{\partial H}{\partial N_k} \!- \!\frac{\partial H}{\partial N_\ell}\Big)\Big] \frac{\partial F}{\partial S}\frac{\partial G}{\partial S}.
\end{aligned}
\end{equation}
One directly checks that the bracket $(F ,G )$ is symmetric, bilinear and satisfies the Leibniz rule, as well as $(H ,S ) = 0 $.  The condition $(S ,S ) \geq 0$ is satisfied if and only if \eqref{thermo_consistency} holds.

In a similar way with the single generator bracket above, this formulation does not impose a specific dependence (such as a linear dependence) of the thermodynamic fluxes $F^{\rm fr}$ and $\mathcal{J}^{\ell\rightarrow k}$ on the thermodynamic forces.
Note that the bracket \eqref{bracked_Kaufman} takes a somehow complicated form. However, as we show below, in the case of a linear relation between the thermodynamic forces and the thermodynamic fluxes, the expression of this bracket is useful to systematically derive a metriplectic bracket.

\subsection{Derivation of the metriplectic bracket}\label{derivation_metripl}

The bracket \eqref{bracked_Kaufman} is not metriplectic since one has
\begin{equation}\label{not_zero}
\begin{aligned}
(F,H)&= \Big\langle  F^{\rm fr}, \frac{\partial F}{\partial p}\Big\rangle\frac{\partial H}{\partial S}
+ \sum_{k<\ell} \mathcal{J}^{\ell\rightarrow k}\Big(\frac{\partial F}{\partial N_k} - \frac{\partial F}{\partial N_\ell}\Big)\frac{\partial H}{\partial S}\neq 0,
\end{aligned}
\end{equation}
in general for an arbitrary function $F$.
Let us assume as in \eqref{thermo_consistency_linear} that the thermodynamic fluxes $F^{\rm fr}$ and $\mathcal{J}^{\ell\rightarrow k}$ depend linearly on their corresponding thermodynamic forces as
\begin{align*}
&F^{\rm fr}\Big(q, \frac{\partial H}{\partial p}, S, N\Big)=- \lambda \cdot \frac{\partial H}{\partial p},\\
&\mathcal{J}^{\ell\rightarrow k}\Big(S, N_k, \frac{\partial H}{\partial N_k},N_\ell, \frac{\partial H}{\partial N_\ell}\Big)= -G^{k\ell}\Big(\frac{\partial H}{\partial N_k} - \frac{\partial H}{\partial N_\ell}\Big),
\end{align*}
where $\lambda=\lambda(q,S):T_qQ\rightarrow T^*_qQ$ is symmetric positive semi-definite and where $G^{k\ell}=G^{k\ell}(S, N_k, N_l)\geq 0$ for all $k,\ell$.
Using these relations in the expression \eqref{not_zero} by writing them in terms of an arbitrary function $G$, and subtracting it from $(F,G)$, we get the symmetric bracket
\begin{align*}
(F,G)_{\rm met}&= (F,G) + \Big\langle  \lambda \cdot \frac{\partial G}{\partial p}, \frac{\partial F}{\partial p}\Big\rangle\frac{\partial H}{\partial S}+ \sum_{k<\ell} G^{k\ell} \Big(\frac{\partial G}{\partial N_k} - \frac{\partial G}{\partial N_\ell}\Big)\Big(\frac{\partial F}{\partial N_k} - \frac{\partial F}{\partial N_\ell}\Big)\frac{\partial H}{\partial S}.
\end{align*}
A direct computation using \eqref{bracked_Kaufman} and rearranging the terms finally yields the expression
\begin{align*}
(F,G)_{\rm met}
&=\frac{1}{\frac{\partial H}{\partial S}}  \left\langle   \frac{\partial F}{\partial p}\frac{\partial H}{\partial S} -  \frac{\partial H}{\partial p}\frac{\partial F}{\partial S}, \lambda\cdot \left( \frac{\partial G}{\partial p}\frac{\partial H}{\partial S} -  \frac{\partial H}{\partial p}\frac{\partial G}{\partial S} \right)\right\rangle\\[3mm]
& + \frac{1}{\frac{\partial H}{\partial S}}\sum_{k<\ell} G^{kl}\Big[\Big(\frac{\partial F}{\partial N_k}-\frac{\partial F}{\partial N_\ell}\Big)\frac{\partial H}{\partial S} - \Big(\frac{\partial H}{\partial N_k} -\frac{\partial H}{\partial N_\ell}\Big)\frac{\partial F}{\partial S}\Big]\\[3mm]
&\qquad \qquad \times\Big[\Big(\frac{\partial G}{\partial N_k} -\frac{\partial G}{\partial N_\ell}\Big)\frac{\partial H}{\partial S}- \Big(\frac{\partial H}{\partial N_k} - \frac{\partial H}{\partial N_\ell}\Big)\frac{\partial G}{\partial S}\Big].
\end{align*}
From this, one directly checks that $(H ,G) _{\rm met}= 0$, and $(G, G)_{\rm met}\geq 0$, for arbitrary $G\in C^\infty(T^*Q\times \mathbb{R}^{K+1})$ by \eqref{thermo_consistency_linear}, therefore $(F,G)_{\rm met}$ is a metriplectic (or GENERIC) bracket. We note that $\dot S=(S,S)=(S,S)_{\rm met}$.
The structure of the first line of the bracket $(\,,)_{\rm met}$ above is a finite dimensional analogue of that of the metriplectic bracket for viscous heat conducting fluid presented in \cite{Mo1984b}. We refer to \cite{ElGB2019} for a similar derivation of the metriplectic bracket for multicomponent fluids, via the variational formulation.

\section{Systems on Lie groups and reduction by symmetries}

We now consider the case where the mechanical configuration space is a Lie group and where both the Lagrangian and the friction force have a symmetry with respect to a subgroup of $G$. We first recall below from \cite{CoGB2019} how the variational formulation  \eqref{LdA_thermo_simple}--\eqref{CV_simple} can be reduced by extending the Euler-Poincar\'e reduction to the case of thermodynamics. From this, the reduced versions of the single and double generator brackets can be derived similarly as above, as well as the metriplectic (or GENERIC) bracket in the case of a linear relation between the forces and the fluxes.
We also establish the relations with the double bracket dissipation developed in \cite{BlKrMaRa1994}. For simplicity, we do not consider the transfer of matter in this section.

\subsection{Variational formulation for thermodynamic systems with symmetries on Lie groups}

Let us assume that $Q=G$ is a Lie group and that the Lagrangian $L:TG \times \mathbb{R}\rightarrow\mathbb{R}$ is left $H$-invariant, where $H\subset G$ is a subgroup. We also assume that the friction force $F^{\rm fr}: TG \times\mathbb{R}\rightarrow T^*G$ is left $H$-equivariant. In local notations, this means
\[
L(hg, hv, S)= L(g,v,S),\qquad F^{\rm fr}(hg, hv, S) = h F^{\rm fr}(g,v,S),
\]
for all $h\in H$. We denote by
\[
N= G/H\ni n= gH
\]
the quotient space. It is naturally acted on by $G$ from the left. For $\xi \in \mathfrak{g}$, the Lie algebra of $G$, we denote by $\xi_N(n)\in T_nN$ the infinitesimal generator of the left action of $G$ on $N$. From the above $H$-invariance, the Lagrangian and the friction force induce their reduced versions
\[
\ell: \mathfrak{g}\times N \times \mathbb{R}\rightarrow \mathbb{R},\qquad f^{\rm fr}: \mathfrak{g}\times N \times \mathbb{R}\rightarrow\mathfrak{g}^*
\]
defined by
\[
L(g,v,S)=\ell( \xi, n, S),\quad F^{\rm fr}(g,v,S)= gf^{\rm fr}(\xi, n,S),
\]
where $\xi=g^{-1}v\in \mathfrak{g}$, $n=g^{-1}H\in N$.

When such symmetries are assumed, the variational formulation  \eqref{LdA_thermo_simple}--\eqref{CV_simple} can be equivalently formulated at the reduced level as follows, see \cite{CoGB2019}.
\medskip
\begin{framed}
\noindent Find the curves $\xi (t)$, $n(t)$, and $S(t)$ which are critical for the \textit{variational condition}
\begin{equation}\label{LdA_thermo_simple_EP}
\begin{aligned}
&\delta \int_{t _1 }^{ t _2}\ell(\xi, n, S){\rm d}t +\int_{t_1}^{t_2} \left<f^{\rm fr}(\xi, n, S), \eta \right>\,{\rm d}t =0,
\end{aligned}
\end{equation}
subject to the \textit{phenomenological constraint}
\begin{equation}\label{CK_simple_EP} 
\frac{\partial \ell}{\partial S}(\xi, n, S)\dot S  =  \left<f^{\rm fr}(\xi, n, S), \xi \right>,
\end{equation}
and for variations subject to the  \textit{variational constraint}
\begin{equation}\label{CV_simple_EP} 
\frac{\partial \ell}{\partial S}(\xi, n, S)\delta S=   \left<f^{\rm ext}(\xi, n, S), \eta \right>,
\end{equation}
and the Euler-Poincar\'e constraints
\begin{equation}\label{EP_constraints} 
\delta\xi= \dot\eta + [\eta, \xi],\quad \delta n + \eta_N(n)=0.
\end{equation}
\end{framed}

This principle yields the following system of evolution equations for the curves $\xi(t)\in \mathfrak{g}$, $n(t)\in N$, $S(t)\in \mathbb{R}$:
\begin{equation}\label{simple_systems_EP}
\left\{
\begin{array}{l}
\displaystyle\vspace{0.2cm} \frac{d}{dt}\frac{\partial\ell}{\partial \xi}= \operatorname{ad}^*_\xi \frac{\partial\ell}{\partial \xi}- \mathbf{J}\Big(\frac{\partial\ell}{\partial n} \Big)+ f^{\rm fr} + f^{\rm ext},\\[3mm]
\displaystyle\frac{\partial\ell}{\partial S}\dot S=   \langle  f^{\rm fr}, \xi  \rangle,\qquad \dot n+ \xi_N(n)=0,
\end{array} \right.
\end{equation}
where the last equation is deduced from the definition $n(t)= g(t)H\in N$ and where $\mathbf{J}: T^\ast N \to \mathfrak{g}^\ast$ is the momentum map, given by 
$\left<\mathbf{J}(\alpha_n),\xi\right>=\left<\alpha_n, \xi_N\right>$ for all $n\in N, \alpha_n \in T^\ast N$ and $\xi \in \mathfrak{g}$.
From now on, we set $ f^{\rm ext}=0$, for simplicity.
In absence of thermodynamic effects, this reduction process recovers the Euler-Poincar\'e reduction, see \cite{HoMaRa1998}, \cite{GBTr2010}.

\subsection{Derivation of the reduced single generator bracket}

Consider the system \eqref{simple_systems_EP}, assume that the Lagrangian $\ell$ is hyperregular and define the associated Hamiltonian $h:\mathfrak{g}^* \times N \times \mathbb{R}\rightarrow\mathbb{R}$ by
\[
h(\mu, n, S)= \langle \mu , \xi \rangle - \ell(\xi, n, S),
\]
where $\xi$ is such that $\frac{\partial \ell}{\partial \xi}=\mu$.
In terms of $h$, system \eqref{simple_systems_matter} can be equivalently written as
\begin{equation}\label{simple_systems_LP}
\hspace{-0.1cm}\left\{
\begin{array}{l}
\displaystyle\vspace{0.2cm} \dot \mu = \operatorname{ad}^*_{\frac{\partial h}{\partial \mu}}\mu + \mathbf{J}\Big(\frac{\partial h}{\partial n} \Big)+ f^{\rm fr},\\[3mm]
\displaystyle - \frac{\partial h}{\partial S}\dot S=   \Big\langle  f^{\rm fr}, \frac{\partial h}{\partial \mu} \Big \rangle,\qquad \dot n+ \Big(\frac{\partial h}{\partial \mu}\Big)_N(n)=0,
\end{array} \right.
\end{equation}
where the dependence of $f^{\rm fr}$ is written in terms of $h$ as
\[
f^{\rm fr}=f^{\rm fr}\Big( \frac{\partial h}{\partial \mu}, n, S\Big).
\]

For a given function $f\in  C^\infty(\mathfrak{g}^*\times N \times \mathbb{R})$, by computing its time derivative
\[
\frac{d}{dt}f= \Big\langle \frac{\partial f}{\partial \mu}, \dot \mu \Big\rangle +  \Big\langle \frac{\partial f}{\partial n}, \dot n\Big\rangle +  \frac{\partial f}{\partial S} \dot S ,
\]
along a solution curve of \eqref{simple_systems_LP}, we directly deduce the single generator form \eqref{single_generator}, with $\{\,,\}$ the Poisson bracket on $\mathfrak{g}^*\times N \times \mathbb{R}$ given by
\begin{equation}\label{LP_bracket}
\begin{aligned}
&\{f,h\}^{\rm red}(\mu,n,S)=- \Big\langle \mu, \Big[\frac{\partial f}{\partial \mu}, \frac{\partial h}{\partial \mu}\Big]\Big\rangle+ \Big \langle\frac{\partial f}{\partial \mu}, \mathbf{J}\Big( \frac{\partial h}{\partial n}\Big)\Big\rangle-\Big \langle\frac{\partial h}{\partial \mu}, \mathbf{J}\Big( \frac{\partial f}{\partial n}\Big)\Big\rangle
\end{aligned}
\end{equation}
and where the dissipation bracket is computed as
\begin{equation}\label{bracked_EB_LP}
\begin{aligned}
& [f ,h]^{\rm red}(\mu,n,S)=\Big\langle  f^{\rm fr}, \frac{\partial f}{\partial \mu}\Big\rangle -\frac{\frac{\partial f}{\partial S}}{\frac{\partial h}{\partial S}}\Big\langle  f^{\rm fr}, \frac{\partial h}{\partial \mu}\Big\rangle.
\end{aligned}
\end{equation}
One directly checks that the conditions $\{h,s\}^{\rm red}=0$, $[h,h]^{\rm red}=0$ are satisfied. The condition $[s,h]^{\rm red}\geq 0$ is satisfied if and only if \eqref{condition_F_fr} holds.

This is the reduced version of the bracket $[\,,]$ given in \eqref{bracked_EB}, in absence of matter transfer.

\subsection{Derivation of the reduced double generator bracket}

Starting again with \eqref{simple_systems_LP} and proceeding exactly as in \S\ref{derivation_double} we get the reduced symmetric bracket
\begin{equation}\label{bracked_Kaufman_LP}
\begin{aligned}
(f,g)^{\rm red}(\mu, n, S)&=\Big\langle  f^{\rm fr}, \frac{\partial f}{\partial \mu}\Big\rangle\frac{\partial g}{\partial S} + \Big\langle  f^{\rm fr}, \frac{\partial g}{\partial \mu}\Big\rangle\frac{\partial f}{\partial S}-\frac{1}{\frac{\partial h}{\partial S}}\Big\langle   f^{\rm fr}, \frac{\partial h}{\partial \mu}\Big\rangle  \frac{\partial f}{\partial S}\frac{\partial g}{\partial S}.
\end{aligned}
\end{equation}
One directly checks that the reduced bracket $(f,g)^{\rm red}$ is symmetric, bilinear and satisfies the Leibniz rule, as well as $(h ,S )^{\rm red} = 0 $.  The condition $(S ,S ) \geq 0$ is satisfied if and only if \eqref{condition_F_fr} holds.

This is the reduced version of the bracket $(\,,)$ given in \eqref{bracked_Kaufman}, in absence of matter transfer.

\subsection{Derivation of the reduced metriplectic bracket}\label{derivation_metriplectic_LP}

Like its unreduced version \eqref{bracked_Kaufman}, the bracket \eqref{bracked_Kaufman_LP} is not metriplectic, since $(f,h)^{\rm red}\neq 0$ in general for an arbitrary function $f$. Let us assume as in \eqref{thermo_consistency_linear} that the friction force $F^{\rm fr}$ depends linearly on the velocity. Its reduced version is
\[
f^{\rm fr}\Big( \frac{\partial h}{\partial\mu},n,S\Big)= - \gamma (n,S)\cdot  \frac{\partial h}{\partial\mu},
\]
where for each $n\in N$ and $S\in \mathbb{R}$, $\gamma(n,S): \mathfrak{g}\rightarrow \mathfrak{g}^*$ is the symmetric positive semi-definite linear map defined from $\lambda(g,S):T_gG\rightarrow T^*_gG$ as
\[
\gamma(n,S) \cdot\xi= g^{-1}\big(\lambda(g,S) \cdot v\big)
\]
with $\xi= g^{-1}v\in\mathfrak{g}$, $n= g^{-1}H\in N$.

Proceeding exactly as in \S\ref{derivation_metripl}, we define the reduced metriplectic bracket from the reduced double generator bracket as
\[
(f,g)^{\rm red}_{\rm met}(\mu, n, S)= (f,g)^{\rm red} + \Big\langle  \gamma \cdot \frac{\partial g}{\partial \mu}, \frac{\partial f}{\partial \mu}\Big\rangle\frac{\partial h}{\partial S}.
\]
From this, a direct computation using \eqref{bracked_Kaufman_LP} and rearranging the terms finally yields the expression
\begin{equation}\label{metriplectic_LP}
\begin{aligned}
&(f,g)_{\rm met}^{\rm red}(\mu, n, S)=\frac{1}{\frac{\partial h}{\partial S}}  \left\langle   \frac{\partial f}{\partial \mu}\frac{\partial h}{\partial S} -  \frac{\partial h}{\partial \mu}\frac{\partial f}{\partial S}, \gamma\cdot \left( \frac{\partial g}{\partial \mu}\frac{\partial h}{\partial S} -  \frac{\partial h}{\partial \mu}\frac{\partial g}{\partial S} \right)\right\rangle.
\end{aligned}
\end{equation}
One directly checks that $(h ,g) ^{\rm red}_{\rm met}= 0$, and $(g, g)^{\rm red}_{\rm met}\geq 0$, for arbitrary $g\in C^\infty(\mathfrak{g}^*\times N \times \mathbb{R})$ since $\gamma$ is positive semi-definite, therefore $(f,g)^{\rm red}_{\rm met}$ is a metriplectic (or GENERIC) bracket. We note that $\dot s=(s,s)^{\rm red}=(s,s)^{\rm red}_{\rm met}$.

\subsection{Coadjoint orbits and double bracket dissipation}

Let us assume that $H=G$, so that system \eqref{simple_systems_LP} reduces to
\begin{equation}\label{simple_systems_LP_G}
\dot \mu = \operatorname{ad}^*_{\frac{\partial h}{\partial \mu}}\mu + f^{\rm fr},\qquad  - \frac{\partial h}{\partial S}\dot S=   \Big\langle  f^{\rm fr}, \frac{\partial h}{\partial \mu} \Big \rangle
\end{equation}
and the variable $n$ is absent.
We note that in general the solutions of this system do not preserve the coadjoint orbits $\mathcal{O}_{\mu_0}= \{\operatorname{Ad}^*_g\mu_0\mid g\in G \}$ of $\mathfrak{g}^*$, which are well-known to be preserved in absence of irreversible processes, \cite{MaRa1999}. Indeed, in this case $f^{\rm fr}=0$ so the first equation in \eqref{simple_systems_LP_G} reduces to the Lie-Poisson equations
\[
\dot \mu = \operatorname{ad}^*_{\frac{\partial h}{\partial \mu}}\mu
\]
on $\mathfrak{g}^*$, while the second gives $S=cst$.

It is however possible to choose the friction force in \eqref{simple_systems_LP_G} in such a way that the coadjoint orbits are preserved. For the development below it is convenient to write the friction force directly in terms of the momentum $\mu$ as $\mathsf{f}^{\rm fr}(\mu, S):= f^{\rm fr}(\frac{\partial h}{\partial\mu}, S)$. Recall that the tangent space to a coadjoint orbit at $\mu\in\mathcal{O}_{\mu_0}$ is $T_\mu \mathcal{O}_{\mu_0}=\{\operatorname{ad}^*_\xi\mu \mid\xi\in \mathfrak{g}\}$, \cite{MaRa1999}. From this expression of the tangent space and from equation \eqref{simple_systems_LP_G}  it is clear that the coadjoint orbits are preserved if and only if the friction force is of the form $\mathsf{f}^{\rm fr}(\mu, S)= \operatorname{ad}^*_{\zeta(\mu,S)}\mu$, for some function $\zeta\in C^\infty(\mathfrak{g}^*\times \mathbb{R})$. From the second law and the second equation in \eqref{simple_systems_LP_G}, the friction force must be dissipative. Since
\[
\Big\langle \mathsf{f}^{\rm fr}(\mu, S), \frac{\partial h}{\partial\mu}\Big\rangle= \Big\langle\! \operatorname{ad}^*_{\zeta(\mu,S)}\mu, \frac{\partial h}{\partial\mu}\Big\rangle =-  \Big\langle \!\operatorname{ad}^*_{\frac{\partial h}{\partial\mu}}\mu, \zeta(\mu,S)\Big\rangle, 
\]
the choice $\zeta (\mu, S)= \big[\operatorname{ad}^*_{\frac{\partial h}{\partial\mu}}\mu\big]^\sharp$, where $\sharp: \mathfrak{g}^*\rightarrow \mathfrak{g}$ is the sharp operator associated to an inner product $\gamma:\mathfrak{g}\times\mathfrak{g}\rightarrow\mathbb{R}$ on $\mathfrak{g}$, yields the dissipative force
\begin{equation}\label{double_bracket_force}
\mathsf{f}^{\rm fr}(\mu, S)= \operatorname{ad}^*_{\big[\operatorname{ad}^*_{\frac{\partial h}{\partial\mu}}\mu\big]^\sharp}\mu.
\end{equation}

In absence of the entropy variable, \eqref{double_bracket_force} recovers the expression of the dissipative external force obtained by {\it double bracket dissipation} in \cite{BlKrMaRa1994}. In our setting, it is interpreted as an internal friction force describing an irreversible process occurring in the system, and leading to an increase
of the entropy.

For the choice \eqref{double_bracket_force}, the reduced single generator bracket \eqref{bracked_EB_LP} becomes
\begin{equation}\label{bracked_EB_LP_double}
\begin{aligned}
& [f ,h]^{\rm red}(\mu, S)=-\gamma\Big( \operatorname{ad}^*_{\frac{\partial f}{\partial \mu}}\mu, \operatorname{ad}^*_{\frac{\partial h}{\partial \mu}}\mu\Big) +\frac{\frac{\partial f}{\partial S}}{\frac{\partial h}{\partial S}}\gamma\Big( \operatorname{ad}^*_{\frac{\partial h}{\partial \mu}}\mu, \operatorname{ad}^*_{\frac{\partial h}{\partial \mu}}\mu\Big).
\end{aligned}
\end{equation}
while the reduced double generator bracket \eqref{bracked_Kaufman_LP} becomes 
\begin{equation}\label{bracked_Kaufman_LP_double}
\begin{aligned}
(f,g)^{\rm red}(\mu, S)&=-\gamma\Big( \operatorname{ad}^*_{\frac{\partial f}{\partial \mu}}\mu, \operatorname{ad}^*_{\frac{\partial h}{\partial \mu}}\mu\Big)\frac{\partial g}{\partial S} -\gamma\Big( \operatorname{ad}^*_{\frac{\partial g}{\partial \mu}}\mu, \operatorname{ad}^*_{\frac{\partial h}{\partial \mu}}\mu\Big)\frac{\partial f}{\partial S}\\[2mm]
&\qquad+\frac{1}{\frac{\partial h}{\partial S}}\gamma\Big( \operatorname{ad}^*_{\frac{\partial h}{\partial \mu}}\mu, \operatorname{ad}^*_{\frac{\partial h}{\partial \mu}}\mu\Big)  \frac{\partial f}{\partial S}\frac{\partial g}{\partial S}.
\end{aligned}
\end{equation}

In order to derive the metriplectic bracket, we shall select a coadjoint orbit $\mathcal{O}_{\mu_0}$ and consider the system \eqref{simple_systems_LP_G} as restricted to $\mathcal{O}_{\mu_0}\times \mathbb{R}$, which is possible with the choice of friction force given in \eqref{double_bracket_force}.

As explained in \cite{BlKrMaRa1994} in absence of entropy variable, when restricted to a given coadjoint orbit $\mathcal{O}_{\mu_0}$, the force \eqref{double_bracket_force} is minus the gradient of the Hamiltonian restricted to $\mathcal{O}_{\mu_0}$, with the gradient computed with the 
respect to the normal metric $\gamma_{\mathcal{O}_{\mu_0}}$ induced on $\mathcal{O}_{\mu_0}$ by the inner product $\gamma$ on $\mathfrak{g}$. In our case, including the entropy variable, we have
\begin{equation}\label{double_bracket_force_gradient}
\mathsf{f}^{\rm fr}(\mu, S)= \operatorname{ad}^*_{\big[\operatorname{ad}^*_{\frac{\partial h}{\partial\mu}}\mu\big]^\sharp}\mu= - \nabla _\mu h(\mu,S)\in T_\mu \mathcal{O}_{\mu_0},
\end{equation}
where, for each fixed $S$, the partial gradient $\nabla _\mu h(\mu,S)\in T_\mu \mathcal{O}_{\mu_0}$ of $h|_{\mathcal{O}_{\mu_0}}$ with respect to $\mu$ is defined by
\[
\gamma_{\mathcal{O}_{\mu_0}} \big( \nabla _\mu h(\mu,S), \delta \mu \big)= \mathbf{d}_\mu (h|_{\mathcal{O}_{\mu_0}})\cdot \delta \mu,\;\forall\; \delta\mu\in T_\mu\mathcal{O}_{\mu_0}.
\]
Here $\mathbf{d}_\mu(h|_{\mathcal{O}_{\mu_0}})\in T^*_\mu\mathcal{O}_{\mu_0}$ denotes the differential of the Hamiltonian $h$ restricted to $\mathcal{O}_{\mu_0}$, the variable $S$ being fixed.

Now using the expression of the friction force given in \eqref{double_bracket_force_gradient} and proceeding similarly as in \S\ref{derivation_metripl} and \S\ref{derivation_metriplectic_LP}, we get the metriplectic bracket on the manifold $\mathcal{O}_{\mu_0}\times \mathbb{R}$ as
\[
\{f,g\}^{\mathcal{O}_{\mu_0}}+ (f,g)^{\mathcal{O}_{\mu_0}}_{\rm met},
\]
where $\{f,g\}^{\mathcal{O}_{\mu_0}}$ is the Poisson bracket associated to the orbit (Kirillov-Kostant-Souriau) symplectic form on $\mathcal{O}_{\mu_0}$ (e.g., \cite{MaRa1999}) and where $(f,g)^{\mathcal{O}_{\mu_0}}_{\rm met}$ is given by
\[
\begin{aligned}
&(f,g)^{\mathcal{O}_{\mu_0}}_{\rm met}(\mu,S)=\frac{1}{\frac{\partial h}{\partial S}} \gamma_{\mathcal{O}_{\mu_0}}\Big(  \nabla_\mu f \frac{\partial h}{\partial S} -  \nabla_\mu h\frac{\partial f}{\partial S}, \nabla_\mu g \frac{\partial h}{\partial S} -  \nabla_\mu h\frac{\partial g}{\partial S}\Big).
\end{aligned}
\]
One directly checks that $(h ,g) ^{\mathcal{O}_{\mu_0}}_{\rm met}(\mu,S)= 0$, and that $(g, g)^{\mathcal{O}_{\mu_0}}_{\rm met}(\mu,S)\geq 0$, for arbitrary $g\in C^\infty(\mathcal{O}_{\mu_0} \times \mathbb{R})$, therefore $(f,g)^{\mathcal{O}_{\mu_0}}_{\rm met}(\mu,S)$ is a metriplectic (or GENERIC) bracket. 

To summarize this paragraph, system \eqref{simple_systems_LP_G} with the friction force chosen as in \eqref{double_bracket_force} preserves the coadjoint orbits and is a thermodynamic extension of the double bracket dissipation equations for Euler-Poincar\'e systems introduced in \cite{BlKrMaRa1994}. Moreover, we have shown that this system can be written by using either the single or the double generator bracket formalism, as well as the metriplectic formalism, when restricted to a coadjoint orbit.


\section{Conclusions}

In this paper we have shown that the variational formulation of nonequilibrium thermodynamics introduced in \cite{GBYo2017a,GBYo2017b} yields a direct and systematic way to derive the main classes of bracket formalisms that have been proposed earlier in the literature. We have illustrated this for the case of a simple system involving a mechanical component together with internal mass transfer and concretely explained how to derive the bracket formalisms in this case. The brackets derived for this case don't seem to have appeared earlier in the literature. We have also shown that reduction by symmetry can be implemented on these bracket formalisms, by using an existing reduction process with irreversible process on the Lagrangian side. From this, we obtained the symmetry reduced versions of the single and double generator brackets, as well as of the metriplectic (or GENERIC) bracket in the case of a linear relation between the forces and the fluxes. We also established the relations with the double bracket dissipation.

While we have considered simple adiabatically closed systems, our approach can be extended to a larger class of systems, such as nonsimple or open systems, following the variational formulation in \cite{GBYo2018a}. We project to explore this issue in a future work, as well as possible relations with the bracket formalism for selective decay developed in \cite{GBHo2013,GBHo2014}.

\paragraph{Acknowledgements.}
F.G.B. is partially supported by the ANR project GEOMFLUID, ANR-14-CE23-0002-01; H.Y. is partially supported by JSPS Grant-in-Aid for Scientific Research (16KT0024, 24224004), the MEXT Top Global University Project and Waseda University (SR 2018K-195, Interdisciplinary institute for thermal energy conversion engineering and mathematics).

                                                   







\end{document}